\documentclass{article}

\usepackage{fullpage} 
\usepackage{amssymb} 
\usepackage[all,dvips]{xy} 
\usepackage{url}
\usepackage{rotating,stmaryrd}

\usepackage{theorem}
\newtheorem{theo}{Theorem}[section]
\newtheorem{prop}[theo]{Proposition}

\theorembodyfont{\normalfont}
\newtheorem{defi}[theo]{Definition}

\newenvironment{proof}{\textit{Proof.} }{$\Box$}

\newcommand{\eps}{\varepsilon} 
\newcommand{\stimes}{\!\times\!} 
\newcommand{\tuple}[1]{\langle#1\rangle} 
\newcommand{\tu}{\langle\,\rangle} 
\newcommand{\frsyn}[2]{{#1}/{#2}} 
\newcommand{\fr}[2]{{#2}\backslash{#1}} 
\newcommand{\cM}{\mathcal{M}} 

\newcommand{\iso}{\cong} 
\newcommand{\To}{\Rightarrow} 
\newcommand{\trnat}{\begin{turn}{-50}\ensuremath{\Uparrow}\end{turn}} 
\newcommand{\rpto}{\rightsquigarrow} 

\newcommand{\bA}{\mathbb{A}} 

\newcommand{\uno}{1}  

\newcommand{\catC}{\mathbf{C}}   
\newcommand{\catS}{\mathbf{S}}   
\newcommand{\catSfr}{\mathbf{S}_\mathbf{2}} 
\newcommand{\catT}{\mathbf{T}}   

\newcommand{\Id}{\mathit{Id}} 
\newcommand{\funY}{\mathcal{Y}} 

\newcommand{\Tt}{\Theta} 
\renewcommand{\tt}{\theta} 
\newcommand{\Ss}{\Sigma} 
\renewcommand{\ss}{\sigma} 
\newcommand{\Set}{\mathit{Set}} 
\newcommand{\Cat}{\mathit{C}} 
\newcommand{\Pro}{\mathit{P}} 
\newcommand{\catD}{\mathit{\Delta}} 

\newcommand{\id}{\mathit{id}} 
\newcommand{\proj}{\mathit{pr}}  

\newcommand{\eq}{\mathit{eq}} 
\newcommand{\dec}{\mathit{dec}} 
\newcommand{\gr}{\mathit{gr}} 
\newcommand{\elem}{\mathit{el}} 
\newcommand{\undec}{\mathit{und}} 
\newcommand{\param}{\mathit{par}} 

\newcommand{\skE}{\mathbf{E}}   
\newcommand{\ske}{\mathbf{e}}   
\newcommand{\Rea}{\mathit{Real}} 

\newcommand{\Type}{\mathtt{Type}}
\newcommand{\Term}{\mathtt{Term}}
\newcommand{\dom}{\mathtt{dom}}
\newcommand{\codom}{\mathtt{codom}}
\newcommand{\selid}{\mathtt{selid}}
\newcommand{\Selid}{\mathtt{Selid}}
\newcommand{\Cons}{\mathtt{Cons}}
\newcommand{\fst}{\mathtt{fst}}
\newcommand{\snd}{\mathtt{snd}}

\newcommand{\comp}{\mathtt{comp}}
\newcommand{\Comp}{\mathtt{Comp}}
\newcommand{\prbone}{\mathtt{b1}}
\newcommand{\prbtwo}{\mathtt{b2}}

\newcommand{\bType}{\mathtt{Type}^\mathtt{2}}

\newcommand{\bCone}{\mathtt{2\texttt{-}Cone}}
\newcommand{\prcone}{\mathtt{c1}}
\newcommand{\prctwo}{\mathtt{c2}}

\newcommand{\bprod}{\mathtt{2\texttt{-}prod}}
\newcommand{\zprod}{\mathtt{0\texttt{-}prod}}
\newcommand{\bProd}{\mathtt{2\texttt{-}Prod}}
\newcommand{\zProd}{\mathtt{0\texttt{-}Prod}}
\newcommand{\btuple}{\mathtt{2\texttt{-}tuple}} 
\newcommand{\ztuple}{\mathtt{0\texttt{-}tuple}} 
\newcommand{\bTuple}{\mathtt{2\texttt{-}Tuple}} 
\newcommand{\zTuple}{\mathtt{0\texttt{-}Tuple}} 
\newcommand{\bbase}{\mathtt{2\texttt{-}base}} 
\newcommand{\zbase}{\mathtt{0\texttt{-}base}}
 
\newcommand{\zbasep}{\mathtt{0\texttt{-}base'}}

\newcommand{\Unit}{\mathtt{Unit}}

\newcommand{\ttid}{\mathtt{id}}
\newcommand{\ttE}{\mathtt{E}}

\newcommand{\deco}{\mathtt{.x}} 
\newcommand{\dotp}{\mathtt{.p}} 
\newcommand{\dotg}{\mathtt{.g}} 
\newcommand{\dotD}{\mathtt{.D}} 
\newcommand{\tti}{\mathtt{i}}
\newcommand{\ttiz}{\mathtt{i0}}
\newcommand{\ttj}{\mathtt{j}}
\newcommand{\ttjz}{\mathtt{j0}}
\newcommand{\ttk}{\mathtt{k}}
\newcommand{\ttkz}{\mathtt{k0}}
\newcommand{\ttc}{\mathtt{c}} 

\renewcommand{\arraystretch}{1.3}  

\title{Diagrammatic logic applied to a parameterization process}  
\author{
C\' esar Dom\'\i nguez \thanks{
Departamento de Matem\'aticas y Computaci\'on,
Universidad de La Rioja,
Edificio Vives, Luis de Ulloa s/n, E-26004 Logro\~no, La Rioja, Spain,
cesar.dominguez@unirioja.es.} 
\and Dominique Duval \thanks{
Laboratoire Jean Kuntzmann, Universit\'e de Grenoble, 
51 rue des math\'ematiques, BP 53, F-38041 Grenoble C\'edex 9, France, 
Dominique.Duval@imag.fr.} 
}
\date{August 25., 2009}

\begin{document}

\maketitle

\begin{itemize}
\item[] \textbf{Abstract.}
This paper provides an abstract definition of some kinds of logics, 
called diagrammatic logics, 
together with a definition of morphisms and of 2-morphisms 
between diagrammatic logics.  
The definition of the 2-category of diagrammatic logics rely on category theory, 
mainly on adjunction, categories of fractions and limit sketches.
This framework is applied to the formalization of a parameterization process.
This process, which consists in adding 
a formal parameter to some operations in
a given specification, is presented as a morphism of logics. 
Then the parameter passing process, 
for recovering a model of the given specification 
from a model of the parameterized specification and an actual parameter,
is seen as a 2-morphism of logics.
\end{itemize}

\section{Introduction}

This paper provides an introduction to the framework of diagrammatic logics
with an application to the formalization of a parameterization process.

The framework of diagrammatic logics is presented in section~\ref{sec:dia}.
It stems from \cite{Du03,Du07}, 
where the aim was to get an abstract definition of logics,
with relevant notions of models and proofs, 
together with a good notion of morphism between logics: 
we were looking for kinds of logics for dealing with computational effects 
and for morphisms for expressing the meaning of the effects into more usual logics. 
This work is based on adjunction \cite{Ka58} and categories of fractions \cite{GZ67} 
with an additional level of abstraction provided by limit sketches \cite{Eh68},
which leads to a notion of entailment apparented to \cite{Makkai97}. 
Our point of view is more abstract than the institutions \cite{GB84},
see \cite{Du03} for a comparison.
This new paper does not depend on \cite{Du03,Du07}.

On the other hand,
the EAT and Kenzo software systems have been developed by F. Sergeraert 
for symbolic computation in algebraic topology \cite{EAT,Kenzo}. 
The data types used in EAT and Kenzo
have been specified through a parameterization process in \cite{DRS06,DLR07},
which is described in \cite{LPR03} in terms of object-oriented technologies
like hidden algebras \cite{GM00} or coalgebras \cite{Ru00}.
The parameterization process consists in adding 
a formal parameter to some operations in a given specification.
It is followed by the parameter passing process, 
which recovers a model of the given specification 
from any model of the parameterized specification and any actual parameter.
A first attempt to use diagrammatic logics in order to formalize 
this parameterization process is given in \cite{DDLR05}. 
In section~\ref{sec:par} we present a simple formalization of 
the parameterization and parameter passing processes
as a morphism and a 2-morphism of diagrammatic logics, respectively.
The focus in this application is on the models, 
but in \cite{DDR09} another kind of application is studied, 
where proofs in a diagrammatic logic play an important role.

Most categorical notions used in this paper can be found in \cite{MacLane98}
or \cite{BW99}. 
For simplicity, we omit most size issues 
and we do not always distinguish between equivalent categories.
The class of morphisms from $X$ to $Y$ in a category $\catC$
is denoted $\catC[X,Y]$.
A \emph{graph} means a directed multigraph,
and in order to distinguish between various kinds of structures 
with an underlying graph 
we speak about the \emph{objects} and \emph{morphisms} of a category, 
the \emph{types} and \emph{terms} of a theory or a specification 
and the \emph{points} and \emph{arrows} of a limit sketch.
The diagrammatic logics which are considered in this paper are the equational logic
and several apparented logics.
However diagrammatic logics can be much richer,
for instance first-order logic as well as simple lambda calculus  
and logics with induction or coinduction can be seen as diagrammatic logics.  

\section{Diagrammatic logics}
\label{sec:dia}

The 2-category of diagrammatic logics and its related notions
are defined in sections~\ref{subsec:dia-ske}, \ref{subsec:dia-log}
and~\ref{subsec:dia-cat}, then the diagrammatic equational logic
is described in section~\ref{subsec:dia-equ}.

\subsection{Limit sketches}
\label{subsec:dia-ske}

There are several definitions of limit sketches (also called projective sketches),
all of them are such that a limit sketch generates a category with limits \cite{CL84,BW99}.
While a category with limits is a graph with identities, compositions, 
limit cones and tuples, satisfying a bunch of axioms, 
we define a \emph{limit sketch} $\skE$ as a graph 
with \emph{potential} identities, compositions, limit cones and tuples,
which become real features in the generated category with limits $\Cat(\skE)$.
For instance a point $X$ in $\skE$ may have a potential identity, 
this is an arrow $\id_X\colon X\to X$ in $\skE$
which becomes the identity morphism at the object $X$ in $\Cat(\skE)$.
As another instance, a diagram in $\skE$ may have a potential limit cone, 
which becomes a limit cone in $\Cat(\skE)$. 
Potential features are not required to satisfy any axiom in $\skE$.
In addition, for  the simplicity of notations,  
we assume that each potential feature is unique:
a point has at most one potential identity,
a diagram has at most one potential limit cone, and so on.

A \emph{morphism} of limit sketches $\ske:\skE_1\to\skE_2$ 
is a graph morphism which maps the potential features of $\skE_1$
to potential features of $\skE_2$.
This forms the category of limit sketches.
A \emph{realization} (or \emph{loose model}) of a limit sketch $\skE$ 
with values in a category $\catC$ 
is a graph morphism which maps the potential features of $\skE$
to real features of $\catC$.
A morphism of realizations is (an obvious generalization of) a natural transformation.
This gives rise to the category $\Rea(\skE,\catC)$
of realizations of $\skE$ with values in $\catC$,
denoted simply $\Rea(\skE)$ when $\catC$ is the category of sets.
The category $\Rea(\skE)$ has colimits
and we will use the fact that left adjoint functors preserve colimits. 

The \emph{Yoneda contravariant realization} $\funY_{\skE}$ of a limit sketch $\skE$
takes its values in $\Rea(\skE)$. 
It is defined as $\funY_{\skE}(E)=\Pro(\skE)[E,-]$
where $\Pro(\skE)$ is the \emph{prototype} of $\skE$,
which means, the category generated by $\skE$
such that every potential feature of $\skE$ becomes a real feature of $\Pro(\skE)$. 
Thanks to $\funY_{\skE}$, up to contravariance 
the limit sketch $\skE$ can be identified to a part of $\Rea(\skE)$
which will be called the \emph{elementary} part of $\Rea(\skE)$
(with respect to $\skE$) and denoted $\Rea_{\elem}(\skE)$.
It is a graph with \emph{distinguished}  features,
defined as the identities, compositions, colimits and cotuples
which are the images of the potential features of $\skE$.
A fundamental property is that the elementary part of $\Rea(\skE)$ 
is \emph{dense} in $\Rea(\skE)$:
every realization or morphism of realizations of $\skE$ 
can be obtained by colimits and cotuples from $\Rea_{\elem}(\skE)$. 
Moeover, a fundamental theorem due to Ehresmann states that 
every morphism of limit sketches  
$\ske\colon \skE_1\to\skE_2$ gives rise to an adjunction $F_{\ske}\dashv G_{\ske}$ 
where the right adjoint $G_{\ske}$ is the precomposition with $\ske$ \cite{Eh68}:
  $$ \xymatrix@C=4pc{ \Rea(\skE_1) \ar[r]^{F_{\ske}}_{\bot} & 
  \Rea(\skE_2) \ar@/^3ex/[l]^{G_{\ske}} } $$
Then the functor $F_{\ske}$ \emph{contravariantly extends} $\ske$
via the Yoneda contravariant realizations,
in the sense that there is a natural isomorphism:
  $$ F_{\ske} \circ \funY_{\skE_1} \iso \funY_{\skE_2} \circ \ske \;.$$

A \emph{locally presentable category} \cite{GU71} is a category $\catC$ 
which is equivalent to the category of set-valued realizations of a limit sketch $\skE$, 
then $\skE$ is called a limit sketch \emph{for} the category $\catC$.
In addition, we define a \emph{locally presentable functor} as 
a functor $F\colon \catC_1\to\catC_2$ which is 
the left adjoint to the precomposition with some morphism of limit sketches $\ske$,
so that $\catC_1$ and $\catC_2$ are locally presentable categories.
Then $\ske$ is called a morphism of limit sketches \emph{for} the functor $F$.

\subsection{Diagrammatic logic: models and proofs}  
\label{subsec:dia-log}

The framework of diagrammatic logics stems from \cite{Du03,Du07}. 

\begin{defi}
\label{defi:dialog}
A \emph{diagrammatic logic} is a locally presentable functor $L$
such that its right adjoint $R$ is full and faithful.
\end{defi}
The fact that $R$ is full and faithful is equivalent to the fact that 
the counit natural transformation $\eps\colon L\circ R\To \Id$ is an isomorphism.
According to \cite{GZ67}, it is also equivalent to the fact that $L$
is a \emph{localization}, up to an equivalence of categories: 
it consists of adding inverse morphisms for some morphisms, 
constraining them to become isomorphisms.
Let us consider a diagrammatic logic $L$:
  $$ \xymatrix@C=4pc{ \catS \ar[r]^{L}_{\bot} & 
  \catT \ar@/^3ex/[l]^{R} } $$
Definition~\ref{defi:dialog} also means that $R$ defines an isomorphim
from $\catT$ to its image, which is a reflective subcategory of $\catS$.

\begin{defi}
The categories $\catS$ and $\catT$ are the category of \emph{specifications}
and the category of \emph{theories}, respectively, of the diagrammatic logic $L$.
A specification $\Ss$ \emph{presents} a theory $\Tt$
if $\Tt$ is isomorphic to $L(\Ss)$.
Two specifications are \emph{equivalent} if they present the same theory. 
\end{defi}
The fact that $R$ is full and faithful means that every theory $\Tt$, 
when seen as a specification $R(\Tt)$, presents itself. 
With the next definition, we claim that every model of a specification
takes its values in some theory.

\begin{defi}
A \emph{(strict) model} $M$ of a specification $\Ss$ in a theory $\Tt$ is 
a morphism of theories $M\colon L\Ss \to \Tt$
or equivalently (thanks to the adjunction) 
a morphism of specifications $M\colon \Ss \to R\Tt$. 
\end{defi}
It follows that equivalent specifications have the same models.  
A model $M$ of $\Ss$ in $\Tt$ is sometimes called an \emph{oblique morphism}, 
it is denoted $M\colon \Ss\to\Tt$.
Whenever in addition $\catS$ and $\catT$ are 2-categories 
with a natural isomorphism between $\catT[L\Ss,\Tt]$ and $\catS[\Ss,R\Tt]$,
then $\catT[L\Ss,\Tt]$ is the \emph{category of models of $\Ss$ in $\Tt$},
denoted $L[\Ss,\Tt]$.
Otherwise, $L[\Ss,\Tt]$ is simply the discrete category with the models 
of $\Ss$ in $\Tt$ as objects.

\begin{defi}
An \emph{entailment} is a morphism $\tau$ in $\catS$ such that 
$L\tau$ is invertible in $\catT$.
\end{defi}
A similar notion can be found in \cite{Makkai97}. 
Two specifications which are related by entailments are equivalent.

\begin{defi}
An \emph{instance} $\rho$ of a specification $\Ss$ in a specification $\Ss_1$
is a cospan in $\catS$ made of a morphism $\ss:\Ss\to\Ss'_1$
and an entailment $\tau:\Ss_1\to\Ss'_1$. 
It is also called a \emph{fraction} with \emph{numerator} $\ss$ and 
\emph{denominator} $\tau$, and it is denoted $\rho=\fr{\ss}{\tau}:\Ss\to\Ss_1$.
\end{defi}
Let us illustrate an instance $\rho=\fr{\ss}{\tau}$ of $\Ss$ in $\Ss_1$ as:
  $$\xymatrix@C=3pc{
     \Ss \ar[r]^{\ss} & \Ss'_1 \ar@<1ex>@{-->}[r] & \Ss_1 \ar[l]^{\tau}   \\
  }$$
this provides easily a diagram in the category $\catS$, by omitting the dotted arrow, 
and a diagram in the category $\catT$, by making the dotted arrow a solid one, 
inverse to $L\tau$:
  $$\mbox{ in }\catS\colon \quad \xymatrix@C=3pc{
     \Ss \ar[r]^{\ss} & \Ss'_1 & \Ss_1 \ar[l]^{\tau}   \\
  } \qquad
   \mbox{ in }\catT\colon \quad \xymatrix@C=3pc{
     L\Ss \ar[r]^{L\ss} & L\Ss'_1 \ar@<1ex>[r]^{(L\tau)^{-1}} & L\Ss_1 \ar[l]^{L\tau}   \\
  }$$
Since the category $\catS$ has colimits 
and since the composition of entailments is an entailment,
the instances can be composed in the usual way as cospans, thanks to pushouts.
This forms the \emph{bicategory of instances} of the logic, denoted $\catSfr$. 
Let $\rho=\fr{\ss}{\tau}\colon \Ss \to \Ss_1$ in $\catSfr$, 
then we define $L\rho=(L\tau)^{-1}\circ L\ss\colon L\Ss \to L\Ss_1$ in $\catT$. 
The instances are better suited than the morphisms of specifications 
for presenting the morphisms of theories, 
because for every morphism of theories $\tt:L\Ss\to L\Ss_1$ 
there is an instance $\rho$ such that $L\rho=\tt$.
Since $L$ is a localization,
the \emph{quotient} category of the bicategory $\catSfr$ 
is equivalent to $\catT$. 

\begin{defi}
An \emph{inference system} for a diagrammatic logic $L$
is a morphism of limit sketches $\ske\colon \skE_S\to\skE_T$ 
for the locally presentable functor $L$. 
\end{defi}
Thanks to the Yoneda contravariant realization, 
the morphism $\ske$ has properties similar to the functor $L$. 
In particular, $\ske$ can be chosen so as to 
consist of adding inverse arrows for some collection of arrows in $\skE_S$;
see \cite[theorem~3.13]{Du03} for a systematic construction of $\ske$.
The next definitions depend on the choice of an inference system 
$\ske\colon \skE_S\to\skE_T$ for $L$; more details are given in \cite{Du07}.

\begin{defi}
An \emph{inference rule} $r$ with \emph{hypothesis} $H$ and \emph{conclusion} $C$
is a span in $\skE_S$, made of two morphisms $t:H'\to H$ and $s:H'\to C$
such that $\ske(t)$ is invertible in $\skE_T$.
It is also called a \emph{fraction} with \emph{numerator} $s$ and 
\emph{denominator} $t$, and it is denoted $r=\frsyn{s}{t}:H\to C$. 
\end{defi}
With this definition we claim that an inference rule 
with hypothesis $H$ and conclusion $C$
can be seen, via the Yoneda contravariant realization, 
as an instance of $\funY(C)$ in $\funY(H)$.
So, we can define an inference step simply as a composition of fractions,
which means, as a pushout in the category $\catS$.

\begin{defi}
Given an inference rule $r=\frsyn{s}{t}:H\to C$
and an instance $\kappa\colon \funY(H) \to \Ss$ 
of the hypothesis $\funY(H)$ in a specification $\Ss$,
the corresponding \emph{inference step} 
provides the instance $\kappa\circ\funY(r)\colon \funY(C) \to \Ss$ 
of the conclusion $\funY(C)$ in $\Ss$.
\end{defi}

\begin{defi}
A \emph{proof} (or \emph{derivation}, or \emph{derived rule}) 
is the description of a fraction in $\catSfr$ in terms of inference rules
(thanks to composition and cotuples).
\end{defi}
Typically, by deriving $\rho=\fr{\id_{\ss}}{\tau}$ for a given morphism $\tau:\Ss_1\to\Ss$,
we get the property that $\tau$ is an entailment.
For instance, in equational logic, let 
$\tau$ be the inclusion of a given specification $\Ss_1$ into the specification 
$\Ss$ made of $\Ss_1$ together with an equation $f=g$ made of two terms $f,g$ in $\Ss_1$;
then $\tau$ is an entailment if and only if the equation $f=g$ holds in 
the theory presented by $\Ss_1$.

\subsection{The 2-category of diagrammatic logics}
\label{subsec:dia-cat}

\begin{defi}
A \emph{morphism of logics} $F\colon L_1\to L_2$ is a pair of locally presentable functors 
$(F_S,F_T)$ together with a natural isomorphism $F_T\circ L_1 \iso L_2\circ F_S$.
\end{defi}
This means that there are inference systems $\ske_1$ and $\ske_2$ 
for $L_1$ and $L_2$ respectively,
and morphisms of limit sketches $\ske_S$ and $\ske_T$  for $F_S$ and $F_T$ 
respectively,
which form a commutative square of limit sketches:
$$  \xymatrix{
L_1 \ar[d]_{F} \\ L_2 \\
} \qquad \qquad \xymatrix@C=3pc{
\catS_1 \ar[r]^{L_1} \ar[d]_{F_S} & \catT_1 \ar[d]^{F_T} \\
\catS_2 \ar[r]^{L_2} & \catT_2 \ar@{}[ul]|{\iso} \\
} \qquad \qquad \xymatrix@C=3pc{
\skE_{1,S} \ar[r]^{\ske_1} \ar[d]_{\ske_S} & \skE_{1,T}  \ar[d]^{\ske_T} \\
\skE_{2,S} \ar[r]^{\ske_2} & \skE_{2,T} \ar@{}[ul]|{=} \\
}$$
Using the Yoneda contravariant realization, 
a morphism of logics $F\colon L_1\to L_2$
can be determined by any graph morphism on $\catS_{1,\elem}$
(the elementary part of $\catS_1$ with respect to $\skE_1$) 
with values in $\catS_2$ preserving the distinguished features 
of $\catS_{1,\elem}$ and the entailments of $L_1$. 
Some morphisms of logics are easier to describe at the sketch level
(as the undecoration morphism in section~\ref{subsec:par-dia})
while others are easier to describe at the logic level
(as the parameterization morphism in section~\ref{subsec:par-par}).
The next result is a straightforward application of adjunction.

\begin{prop}
\label{prop:mod}
Given a morphism of logics $F\colon L_1\to L_2$ and 
the corresponding adjunctions 
$F_T \dashv G_T$ between theories 
and $F_S \dashv G_S$ between specifications, 
for each specification $\Ss_1$ of $L_1$ and each theory $\Tt_2$ of $L_2$ 
the adjunctions provide an isomorphism, natural in $\Ss_1$ and $\Tt_2$,
between the categories of models:
  $$ L_1[\Ss_1,G_T(\Tt_2)] \iso L_2[F_S(\Ss_1),\Tt_2] \;. $$
\end{prop}

\begin{defi}
A \emph{2-morphism of logics} $\ell\colon F\To F'\colon L_1\to L_2$ 
is a pair of natural transformations $(\ell_S,\ell_T)$ 
where $\ell_S\colon F_S\To F'_S\colon \catS_1\to \catS_2$ 
and $\ell_T\colon F_T\To F'_T\colon \catT_1\to \catT_2$ 
are such that $\ell_T\circ L_1 = L_2\circ \ell_S$. 
\end{defi}
Given a morphism of logics $F=(F_S,F_T)$ or a 2-morphism of logics $\ell=(\ell_S,\ell_T)$,  
we will usually omit the subscripts $S$ and $T$.

The diagrammatic logics together with their morphisms and 2-morphisms form 
a 2-category. By focusing on theories we get a functor 
from the 2-category of diagrammatic logics to the 2-category of categories. 
The other parts of the logic (the category of specifications, the adjunction, 
and the inference system) provide a way to answer some issues about theories,
typically whether some morphisms of theories are invertible. 

\subsection{The diagrammatic equational logic}
\label{subsec:dia-equ}

The equational logic provides a fundamental example of a diagrammatic logic.
As usual in categorical logic (see \cite{Pitts}), 
the \emph{equational theories} are defined as the categories with chosen finite products;
with the functors which preserve the chosen finite products
they form a category $\catT_{\eq}$.
Similarly (see \cite{Lellahi89,BW99,Wells93}), 
the \emph{equational specifications} are defined as the finite product sketches,
which means, the limit sketches (as in section~\ref{subsec:dia-ske}) 
such that their potential limits are only potential products; 
with the morphisms of finite product sketches 
they form a category $\catS_{\eq}$.
Since all finite products may be recovered from binary products and a terminal type,
we restrict the arity of products to either~2 or~0.
We will often omit the word ``equational''. 
Every theory $\Tt$ can be seen as a specification $R_{\eq}\Tt$ 
and every specification $\Ss$ generates, or presents, a theory $L_{\eq}\Ss$.
This corresponds to an adjunction:
  $$ \xymatrix@C=4pc{ \catS_{\eq} \ar[r]^{L_{\eq}}_{\bot} & 
  \catT_{\eq} \ar@/^3ex/[l]^{R_{\eq}} } $$
The category of sets with the cartesian products as chosen products forms 
an equational theory denoted $\Set$.
By default the models of an equational specification $\Ss$
are the models of $\Ss$ in $\Set$, called the \emph{set-valued models} of $\Ss$. 
It is a classical exercise to build limit sketches for $\catT_{\eq}$ and $\catS_{\eq}$, 
then it is easy to check that $L_{\eq}$ is a diagrammatic logic.
A simplified description is given now, see \cite{DD09} for a detailed construction.
The starting point is the limit sketch for graphs $\skE_{\gr}$,
where the points $\Type$ and $\Term$ stand for 
the sets of vertices (or types) and edges (or terms)
and the arrows $\dom$ and $\codom$ for the functions 
source (or domain) and target (or codomain):
$$
  \xymatrix@C=4pc{
    \;\Type\; & \;\Term\; \ar[l]_{\dom} \ar@<1ex>[l]^{\codom} \\
    }
$$
Figure~\ref{fig:skEeqS} presents the main part of the graph underlying $\skE_{\eq,S}$,
in addition there are potential limits, including the specification of potential monomorphisms, 
and equalities of arrows.
We have represented this graph in such a way that the bottom line,
which is made of $\skE_{\gr}$ with potential limits and tuples, 
is equivalent to $\skE_{\gr}$. 
The point $\Type$ has been duplicated for readability,
and the point $\Unit$ is a potential terminal type, interpreted as a singleton.

\begin{itemize}
\item The point $\Comp$ stands for the set of pairs of composable terms,
the arrow $\tti$ for the inclusion into the set of pairs of consecutive terms
and $\comp$ for $(f,g)\mapsto g\circ f$. 
\item The point $\Selid$ stands for the set of types with a potential identity,
the arrow $\ttiz$ for the inclusion and $\selid$ for $X\mapsto \id_X$.
\item The point $\bProd$ stands for the set of pairs of types with a potential binary product,
the arrow $\ttj$ for the inclusion into the set of pairs of types
and $\bprod$ for $(Y_1,Y_2)\mapsto (\proj_i\colon Y_1\times Y_2\to Y_i)_{i=1,2}$. 
\item The point $\bTuple$ stands for the set of binary cones with a potential binary tuple, 
the arrow $\ttk$ for the inclusion into the set of binary cones, 
$\bbase'$ for recovering the base $(f_i\colon X\to Y_i)_{i=1,2}\mapsto (Y_1,Y_2)$, 
and $\btuple$ stands for the construction of the potential binary tuple
$(f_i\colon X\to Y_i)_{i=1,2}\mapsto \tuple{f_1,f_2}\colon X\to Y_1\times Y_2$.
\item The point $\zProd$ stands for the set of potential terminal types,
the arrow $\ttjz$ for the injection 
(ensuring that there is at most one terminal type) 
and $\zprod$ for the selection of the potential terminal type (if any).
\item The point $\zTuple$ stands for the set of types with a potential 
collapsing term (or nullary tuple), 
the arrow $\ttkz$ for the inclusion into the set of types, 
$\zbasep$ for recovering the potential terminal type
and $\ztuple$ stands for the construction of the potential collapsing term
$X\mapsto \tu_X\colon X\to \uno$.
\end{itemize}

\begin{figure}[ht]
$$
 \xymatrix@C=3pc{
      \zProd \ar[d]_{\ttjz} \ar@<1ex>[rd]|{\zprod} & 
          \zTuple \ar[l]_{\zbasep} \ar[d]^{\ttkz}  \ar@/^7ex/@<1ex>[rrd]^(.4){\ztuple} &
        \Selid \ar[d]_{\ttiz} \ar@<1ex>[rd]|{\selid} & &
       \Comp \ar[d]^{\tti} \ar@<-1ex>[dl]|{\comp}
      & \bTuple\ar[r]^{\bbase'}\ar[d]_{\ttk}\ar@/_7ex/@<-1ex>[lld]_(.4){\btuple}& 
          \bProd \ar[d]^{\ttj}  \ar@<-1ex>[ld]|{\bprod} \\
   \Unit & \Type \ar[l]_{\zbase} \ar[r]^{\ttid} & \Type & 
      \Term \ar@<1ex>[l]^{\codom} \ar[l]_{\dom} &
      \Cons \ar@<1ex>[l]^{\snd} \ar[l]_{\fst} &
      \bCone  \ar@/^5ex/@<1ex>[ll]^(.4){\prctwo} \ar@/^5ex/[ll]_(.4){\prcone} \ar[r]^{\bbase} &
      \bType \ar@/^10ex/@<1ex>[llll]^(.3){\prbtwo} \ar@/^10ex/[llll]_(.3){\prbone} \\
    }
$$
\caption{\label{fig:skEeqS} The graph underlying $\skE_{\eq,S}$}
\end{figure}

A limit sketch $\skE_{\eq,T}$ for equational theories is obtained from $\skE_{\eq,S}$ 
by choosing the entailments and mapping them to equalities, 
the corresponding morphism is the \emph{diagrammatic equational logic} $L_{\eq}$.
Figure~\ref{fig:equ} provides the correspondence between 
the usual rules of equational logic and the diagrammatic inference rules, as fractions.
Since only a part of $\skE_{\eq,S}$ is considered, some rules are missing, 
it is an exercise to enlarge $\skE_{\eq,S}$ so as to get them.

\renewcommand{\arraystretch}{2} 
\begin{figure}[ht]
$$\begin{array}{|l|c|c|}
\hline
\multicolumn{1}{|c|}{\textrm{name}} & \textrm{rule} & \textrm{fraction} \\ 
\hline
\hline
  \textrm{composition} &
    \frac{f\colon X\to Y \quad g\colon Y\to Z}{g\circ f\colon X\to Z} &
    $\xymatrix@C=3pc{
      \Cons \ar@<1ex>@{-->}[r] & \Comp \ar[l]^{\tti} \ar[r]^{\comp} & \Term \\
    }$ \\  
\hline
  \textrm{identity} &
    \frac{X}{\id_X\colon X\to X} &
  $\xymatrix@C=3pc{
     \Type \ar@<1ex>@{-->}[r] & \Selid \ar[l]^{\ttiz} \ar[r]^{\selid} & \Term \\
  }$ \\
\hline
  \textrm{binary product} &
     \frac{Y_1 \quad Y_2}
    { \proj_i\colon Y_1\times Y_2\to Y_i}_{i=1,2} &
  $\xymatrix@C=3pc{
     \bType \ar@<1ex>@{-->}[r] & \bProd \ar[l]^{\ttj} \ar[r]^{\bprod} & \bCone  \\
  }$ \\
\hline
  \textrm{binary tuple} &  
     \frac{f_1\colon X\to Y_1 \; f_2\colon X\to Y_2}
     {\tuple{f_1,f_2} \colon X\to Y_1\times Y_2}  &
  $\xymatrix@C=3pc{
     \bCone \ar@<1ex>@{-->}[r] & \bTuple \ar[l]^{\ttk} \ar[r]^{\btuple} & \Term \\
  }$ \\
\hline
  \textrm{terminal type} &
    \frac{}{\;\uno\;} &
$\xymatrix@C=3pc{
     \Unit \ar@<1ex>@{-->}[r] & \zProd \ar[l]^{\ttjz}  \ar[r]^{\zprod} & \Type \\
  }$ \\
\hline
  \textrm{collapsing} &
    \frac{X}{\tu_X\colon X\to \uno} &
$\xymatrix@C=3pc{
     \Type \ar@<1ex>@{-->}[r] & \zTuple \ar[l]^{\ttkz}  \ar[r]^{\ztuple} & \Term \\ 
  }$ \\
\hline
\end{array}$$
\caption{\label{fig:equ} Rules for the equational logic}
\end{figure}
\renewcommand{\arraystretch}{1.3} 

It should be noted that in this definition of the equational 
theories and specifications, the equations are identities of terms; 
a more subtle point of view, where the equations in a theory form a congruence, 
can be found in \cite{DD09}. 

\section{A parameterization process}
\label{sec:par}

Several variants of the diagrammatic equational logic,
related by morphisms, are defined in section~\ref{subsec:par-dia}.
The parameterization process and the parameter passing process 
are formalized in sections~\ref{subsec:par-par} and~\ref{subsec:par-pas}, respectively.

\subsection{Some diagrammatic logics}
\label{subsec:par-dia}

The theories of the \emph{parameterized equational logic} $L_A$ 
are the equational theories together with a distinguished type, 
called the \emph{type of parameters} and usually denoted $A$. 
The specifications are the equational specifications 
with maybe a distinguished type $A$. 
The inclusion of limit sketches determines a morphism of logics 
$ F_A\colon L_{\eq} \to L_A$. 

The theories of the \emph{equational logic with a parameter} $L_a$ 
are the parameterized equational theories 
together with a distinguished constant of type $A$,
called the \emph{parameter} and usually denoted $a\colon \uno\to A$.
The specifications are the parameterized equational specifications 
with maybe a distinguished term $a\colon \uno\to A$.
The inclusion of limit sketches determines a morphism of logics 
$ F_a\colon L_A \to L_a$.

The theories of the \emph{decorated equational logic} $L_{\dec}$ 
are the equational theories 
together with a wide subtheory called \emph{pure} 
(\emph{wide} means with the same types). 
The specifications are the equational specifications 
together with a wide subspecification. 
Here is a way to build $\skE_{\dec,T}$ from $\skE_{\eq,T}$
which reflects the meaning of the word ``decoration'',
a smaller choice for $\skE_{\dec,T}$ can be found in \cite{DD09}.
The decorations in this context are simply made of two keywords 
$p$ for ``pure'' and $g$ for ``general'';
some terms are pure, all terms are general, and there are rules for 
dealing with the decorations: 
identities and projections are always pure,
and the compositions or tuples of pure terms are pure.
This information can be encoded as a realization $\catD$ of $\skE_{\eq,T}$ 
with values in the category of equational theories, as follows.
First let us describe the set-valued realization $\catD_0$ of $\skE_{\eq,T}$ 
underlying $\catD$.
The set $\catD_0(\Type)$ is made of one type $D$
and the set $\catD_0(\Term)$ of two terms $p$ and $g$,
so that $\catD_0(\Cons)=\{(p,p),(p,g),(g,p),(g,g)\}$,
$\catD_0(\bCone)=\{(p,p),(p,g),(g,p),(g,g)\}$
and $\catD_0(\bType)=\{(D,D)\}$, 
and we denote $\catD_0(\Unit)=\{\star\}$.
Then $\catD_0(\selid)$ maps $D$ to $p$,  
$\catD_0(\comp)$ maps $(p,p)$ to $p$ and everything else to $g$, 
$\catD_0(\bprod)$ maps $(D,D)$ to $(p,p)$, 
$\catD_0(\btuple)$ maps $(p,p)$ to $p$ and everything else to $g$, 
$\catD_0(\zprod)$ maps $\star$ to $p$ 
and $\catD_0(\ztuple)$ maps $D$ to $p$.
The structure of equational theory on each set $\catD_0(E)$ is induced by 
a monomorphism $p\to g$ in $\catD(\Term)$.
Then $\skE_{\dec,T}$ is the \emph{sketch of elements} 
(similar to the more usual \emph{category of elements}) 
of the realization $\catD$ of $\skE_{\eq,T}$:
the points of $\skE_{\dec,T}$ include
one point $\Type\dotD$ over the point $\Type$ of $\skE_{\eq,T}$,
two points $\Term\dotp$ and $\Term\dotg$ over the point $\Term$ of $\skE_{\eq,T}$,
four points over $\Cons$, and so on,
and the arrows of $\skE_{\dec,T}$ include
an arrow $\ttc\colon \Term\dotp \to \Term\dotg$ over $\ttid_{\Term}$ 
which is a potential monomorphism,
for the conversion of pure terms to general terms.

Clearly by forgotting the decorations we get 
a morphism of diagrammatic logics $ F_{\undec}\colon L_{\dec} \to L_{\eq}$,
called the \emph{undecoration} morphism. 
And by mapping every feature of $\skE_{\eq,T}$
to the corresponding pure feature of $\skE_{\dec,T}$ we get 
a morphism of diagrammatic logics $F_p\colon L_{\eq} \to L_{\dec}$ 
such that $F_{\undec}\circ F_p = \id_{L_{\eq}}$. 

\subsection{The parameterization process is a morphism of logics}
\label{subsec:par-par}

In this section we define a morphism of logics 
$ F_{\param}\colon L_{\dec} \to L_A $.
We define $F_{\param}$ on specifications, its definition on theories follows easily.
We will use the fact, which follows from the definition of a morphism of logics, 
that a specification may be replaced by an equivalent one whenever needed. 

The parameterization process starts from a decorated specification 
and returns a parameterized specification. 
Roughly speaking, it replaces every 
general feature in a decorated specification by a parameterized one, 
in such a way that a pure feature does not really depend on the parameter.
More precisely, types and pure terms are unchanged,
while every general term $f\colon X\to Y$ is replaced by $f'\colon A\times X\to Y$ 
where $A$ is the type of parameter. 
Figure~\ref{fig:F-elem} defines the image of the elementary decorated specifications 
(pure terms are denoted with ``$\rpto$''
and the projections $\proj_X\colon A\times X\to A$
and $\eps_X\colon A\times X\to X$ are often omitted): 
for each point $\ttE\deco$ in $\skE_{\dec,S}$,
the parameterization process replaces 
the elementary decorated specification $ \funY(\ttE\deco)$
by the parameterized specification $F_{\param}(\funY(\ttE\deco))$.
The morphisms between elementary decorated specifications
are transformed in a straightforward way. 
For instance, the image of the morphism $\funY(\ttc)$, 
where $\ttc\colon \Term\dotp \to \Term\dotg$ is the conversion arrow,
maps $f'\colon A\times X\to X$ in $F_{\param}(\funY(\Term\dotg))$
to $f\circ \eps_X \colon A\times X\to Y$ in $F_{\param}(\funY( \Term\dotp))$,
or more precisely in a parameterized specification equivalent to 
$F_{\param}(\funY( \Term\dotp))$.
This provides a graph morphism 
$ F_{\param}\colon \Rea_{\elem}(\skE_{\dec,S})\to\Rea(\skE_{A,S})$. 

\begin{figure}[ht]
$$ \begin{array}{|l|l|c|c|}
\hline
 & \textrm{point} \; \ttE\deco & \funY(\ttE\deco) & F_{\param}(\funY(\ttE\deco)) \\
\hline
\hline
  \textrm{type} & \Type\dotp & X & X \\ 
\hline
  \textrm{pure term} & \Term\dotp & 
  \xymatrix@C=3pc{X\ar@{~>}[r]^{f} & Y} & 
  \xymatrix@C=3pc{X\ar[r]^{f} & Y} \\ 
\hline
  \textrm{term} & \Term\dotg &
  \xymatrix@C=3pc{X\ar[r]^{f} & Y} & 
  \xymatrix@C=3pc{A\stimes X\ar[r]^{f'} & Y}  \\ 
\hline
  \textrm{pure composition} & \Comp\dotp & 
  \xymatrix@C=3pc{X\ar@{~>}[r]^{f} \ar@/_3ex/@{~>}[rr]_{g\circ f}^{=} & Y\ar@{~>}[r]^{g} & Z} & 
  \xymatrix@C=3pc{X\ar[r]^{f} \ar@/_3ex/[rr]_{g\circ f}^{=} & Y\ar[r]^{g} & Z} \\
\hline
  \textrm{composition} & \Comp\dotg &  
  \xymatrix@C=3pc{X\ar[r]^{f} \ar@/_3ex/[rr]_{g\circ f}^{=} & Y\ar[r]^{g} & Z} & 
  \xymatrix@C=3pc{A\stimes X \ar[r]^{\tuple{\proj_X,f'}} 
     \ar@/_3ex/[rr]_{g'\circ\tuple{\proj_X,f'}}^{=} & A\stimes Y\ar[r]^{g'} & Z \\} \\ 
\hline
  \textrm{selection of identity} & \Selid\dotp &
  \xymatrix@C=3pc{X\ar@{~>}[r]^{\id_X} & X} & 
  \xymatrix@C=3pc{X\ar[r]^{\id_X} & X} \\
\hline
  \textrm{binary product} & \bProd\dotp &  
  \xymatrix@C=3pc@R=.2pc{ Y_1 & \\ 
    & Y_1\stimes Y_2 \ar@{~>}[lu]_{p_1} \ar@{~>}[ld]^{p_2} \\ Y_2 & \\ }  & 
  \xymatrix@C=3pc@R=.2pc{ Y_1 & \\ 
    & Y_1\stimes Y_2 \ar[lu]_{p_1} \ar[ld]^{p_2} \\ Y_2 & \\ } \\ 
\hline
  \textrm{pure pairing} & \bTuple\dotp & 
  \xymatrix@C=3pc@R=.5pc{ & Y_1 & \\ 
    X \ar@{~>}[ru]^{f} \ar@{~>}[rd]_{g} \ar@{~>}[rr]|{\,\tuple{f,g}\,} & \ar@{}[u]|{=}\ar@{}[d]|{=} & 
    Y_1\stimes Y_2 \ar@{~>}[lu]_{p_1} \ar@{~>}[ld]^{p_2} \\ & Y_2 & \\ }  & 
  \xymatrix@C=3pc@R=.5pc{ & Y_1 & \\ 
    X \ar[ru]^{f} \ar[rd]_{g} \ar[rr]|{\,\tuple{f,g}\,} & 
    \ar@{}[u]|{=}\ar@{}[d]|{=} & 
    Y_1\stimes Y_2 \ar[lu]_{p_1} \ar[ld]^{p_2} \\ & Y_2 & \\ } \\ 
\hline
  \textrm{pairing} & \bTuple\dotg & 
  \xymatrix@C=3pc@R=.5pc{ & Y_1 & \\ 
    X \ar[ru]^{f} \ar[rd]_{g} \ar[rr]|{\,\tuple{f,g}\,} & \ar@{}[u]|{=}\ar@{}[d]|{=} & 
    Y_1\stimes Y_2 \ar@{~>}[lu]_{p_1} \ar@{~>}[ld]^{p_2} \\ & Y_2 & \\ }  & 
  \xymatrix@C=3pc@R=.5pc{ & Y_1 & \\ 
    A\stimes X \ar[ru]^{f'} \ar[rd]_{g'} \ar[rr]|{\,\tuple{f',g'}\,} & 
    \ar@{}[u]|{=}\ar@{}[d]|{=} & 
    Y_1\stimes Y_2 \ar[lu]_{p_1} \ar[ld]^{p_2} \\ & Y_2 & \\ } \\ 
\hline
  \textrm{terminal type} & \zProd\dotp &  
  \xymatrix@C=3pc@R=.7pc{ \uno \\ }  & 
  \xymatrix@C=3pc@R=.7pc{ \uno \\ } \\ 
\hline
  \textrm{pure collapsing} & \zTuple\dotp & 
  \xymatrix@C=3pc@R=1pc{ 
    X \ar@{~>}[r]^{\tu_X} & \uno  \\ }  & 
  \xymatrix@C=3pc@R=1pc{ 
    X \ar[r]^{\tu_X} & \uno  \\ }  \\  
\hline
\end{array}$$
\caption{\label{fig:F-elem} The parameterization morphism 
on elementary decorated specifications}
\end{figure}

\begin{theo}
\label{theo:param}
The graph morphism $F_{\param}$ 
defines a morphism of diagrammatic logics: 
  $$ F_{\param}\colon L_{\dec} \to L_A $$ 
which is the inclusion on the pure part of $L_{\dec}$,
in the sense that $F_{\param}\circ F_p = F_A$.
It is called the \emph{parameterization} morphism.
\end{theo}

\begin{proof}
It can be checked that this graph morphism 
preserves the distinguished features of $\Rea_{\elem}(\skE_{\dec,S})$
and the entailments of the decorated logic, 
so that it provides a morphism of diagrammatic logics.
The equality $F_{\param}\circ F_p = F_A$ 
is easily checked on elementary specifications.
\end{proof}

The morphisms of logics $F_{\undec}$, $F_{\param}$ and $F_A$
form a (non-commutative) triangle, which becomes commutative when restricted to 
the pure part of $L_{\dec}$:
$$ \xymatrix@C=3pc@R=1pc { 
& L_{\eq} \ar[d]_{F_p} \ar@/_2ex/[ldd]_(.4){\id}^(.4){=} \ar@/^2ex/[rdd]^(.4){F_A}_(.4){=} & \\
& L_{\dec} \ar[ld]_(.4){F_{\undec}} \ar[rd]^(.4){F_{\param}} &  \\
L_{\eq} \ar[rr]_{F_A} & & L_A \\
} $$
The parameterization morphism $F_{\param}$ formalizes 
the parameterization process. 
The span made of $F_{\undec}$ and $F_{\param}$ formalizes 
the process of starting from an equational specification $\Ss_{\eq}$,
choosing a pure subspecification $\Ss_0$ of $\Ss_{\eq}$ 
so as to get a decorated specification $\Ss_{\dec}$ such that 
$\Ss_{\eq}=F_{\undec}(\Ss_{\dec})$,
then forming the parameterized specification $\Ss_A=F_{\param}(\Ss_{\dec})$.

\subsection{The parameter passing process is a 2-morphism of logics}
\label{subsec:par-pas}

The diagram of logics in section~\ref{subsec:par-par} 
composed with the inclusion $F_a\colon L_A\to L_a$, 
which adds the parameter $a\colon \uno\to A$, provides another diagram 
with in addition a 2-morphism $\ell$ as described below:
$$ \xymatrix@C=3pc@R=1pc { 
& L_{\eq} \ar[d]_{F_p} \ar@/_2ex/[ldd]_(.4){\id}^(.4){=} \ar@/^2ex/[rdd]^(.4){F_a\circ F_A}_(.4){=} & \\
& L_{\dec} \ar[ld]_(.4){F_{\undec}} \ar[rd]^(.4){F_a\circ F_{\param}\quad} &  \\
L_{\eq} \ar[rr]_{F_a\circ F_A} \ar@{}[rru]|{\trnat}_{\ell} & & L_a \\
} $$
Each decorated specification $\Ss_{\dec}$, with $\Ss_{\eq}=F_{\undec}(\Ss_{\dec})$,
gives rise to two specifications with parameter:  
on the one hand $\Ss_{\eq,a}=F_a(F_A(\Ss_{\eq}))$, 
which is simply $\Ss_{\eq}$ seen as a specification with a parameter, 
and on the other hand $\Ss_a=F_a(F_{\param}(\Ss_{\dec}))$.
Let us define the morphism $\ell_{\Ss_{\dec}}\colon  \Ss_{\eq,a} \to \Ss_a$.
When $\Ss_{\dec}$ is some $\funY(\ttE\dotp)$ (where $\mathtt{p}$ means ``pure'') 
it is easy to check that $\Ss_{\eq,a}=\Ss_a$; then $\ell_{\Ss_{\dec}}$ is the identity. 
When $\Ss_{\dec}=\funY_{\dec}(\Term\dotg)$ (where $\mathtt{g}$ means ``general'') ,
then $\ell_{\Ss_{\dec}}$ is defined by 
$\ell_{\Ss_{\dec}}(f)=f'\circ\tuple{a,\id_X}\colon X\to Y$
(where $\uno\times X$ is identified with $X$).
The definitions when $\Ss_{\dec}=\funY_{\dec}(\Comp\dotg)$
and when $\Ss_{\dec}=\funY_{\dec}(\bTuple\dotg)$ are similar.

\begin{theo}
\label{theo:passing}
The morphisms $\ell_{\Ss_{\dec}}\colon  \Ss_{\eq,a} \to \Ss_a$
define a 2-morphism of diagrammatic logics:
 $$ \ell \colon F_a\circ F_A\circ F_{\undec} \To F_a\circ F_{\param}
   \colon L_{\dec}\to L_a $$
which is the identity on the pure part of $L_{\dec}$. 
It is called the \emph{parameter passing} 2-morphism.
\end{theo}

\begin{proof}
The definition of $\ell_{\Ss_{\dec}}$ on the elementary decorated specifications 
is extended to all specifications by colimits, and the result follows. 
\end{proof}

Theorem~\ref{theo:passing} has the expected consequence on models,
stated as proposition~\ref{prop:passing}:
given a set-valued model $M_A$ of the paramererized specification $\Ss_A$, 
each $\alpha\in M_A(A)$, called an \emph{actual parameter} or an \emph{argument},
gives rise to a model $\cM(\alpha)$ of the equational specification $\Ss_{\eq}$.
Let us introduce some notations.
For each set $\bA$, let $\Set_{\bA}$ denote the object of $\catT_A$
made of the equational theory of sets with $\bA$ 
as the interpretation of $A$, so that $R_A(\Set_{\bA})=\Set$.
For each set $\bA$ and element $\alpha\in\bA$,
let $\Set_{\bA,\alpha}$ denote the object of $\catT_a$
made of the equational theory of sets with $\bA$ and $\alpha$
as the interpretations of $A$ and $a$ respectively, 
so that $R_a(\Set_{\bA,\alpha})=\Set_{\bA}$.
For each decorated specification $\Ss_{\dec}=(\Ss_{\eq},\Ss_0)$, 
made of an equational specification $\Ss_{\eq}$ and a wide subspecification $\Ss_0$,
and for each set-valued equational model $M_0$ of $\Ss_0$,
let $L_{\eq}[\Ss_{\eq},\Set]|_{M_0}$ denote the set of models of $\Ss_{\eq}$ 
extending $M_0$.
Let $\Ss_A=F_{\param}(\Ss_{\dec})$,
the definition of $F_{\param}$ is such that $\Ss_0$ is also a subspecification of $\Ss_A$
and for each $f\colon X\to Y$ in $\Ss_{\eq}$
there is a $f'\colon A\times X\to Y$ in $\Ss_A$, with $f'=f\circ\eps_X$ when $f$ is pure.

\begin{prop}
\label{prop:passing}
Let $\Ss_{\dec}=(\Ss_{\eq},\Ss_0)$ be a decorated specification
and let $\Ss_A=F_{\param}(\Ss_{\dec})$.
For each set $\bA$ and each set-valued model $M_A\colon \Ss_A\to\Set_{\bA}$ in $L_A$,
let $M_0\colon \Ss_{\eq}\to\Set$ denote the restriction of $M_A$ to $\Ss_0$.
Then there is a function: 
  $$ \cM \colon  \bA \to L_{\eq}[\Ss_{\eq},\Set]|_{M_0} $$ 
which maps each $\alpha\in\bA$ to the model $\cM(\alpha)$ of $\Ss_{\eq}$ extending $M_0$ 
and such that $\cM(\alpha)(f)=M_A(f')(\alpha,-)$ for each $f\colon X\to Y$ in $\Ss_{\eq}$.
\end{prop}

\begin{proof}
Let $\Ss_{\eq,a}=F_a(F_A(\Ss_{\eq}))$ and $\Ss_a=F_a(F_{\param}(\Ss{\dec}))$.
The precomposition with the morphism $\ell_{\Ss_{\dec}}\colon \Ss_{\eq,a} \to \Ss_a$ 
gives rise to a functor $L_a[\Ss_a,\Set_{\bA,\alpha}] \to L_a[\Ss_{\eq,a},\Set_{\bA,\alpha}]$.
Proposition~\ref{prop:mod} provides the isomorphisms 
$L_a[\Ss_a,\Set_{\bA,\alpha}] \iso L_A[\Ss_A,\Set_{\bA}]$
and $L_a[\Ss_{\eq,a},\Set_{\bA,\alpha}] \iso L_{\eq}[\Ss_{\eq},\Set]$.
So, for each $\alpha\in\bA$ we get a functor 
$ L_A[\Ss_A,\Set_{\bA}] \to L_{\eq}[\Ss_{\eq},\Set] $.
Let  $M_{A,\alpha}$ denote the image of $M_A$,
because of the definition of $\ell_{\Ss_{\dec}}$
it extends $M_0$ and satisfies $M_{A,\alpha}(f)=M_A(f')(\alpha,-)$ 
for each $f\colon X\to Y$ in $\Ss_{\eq}$.
Now, when $M_A$ is fixed, the result follows 
by defining $\cM(\alpha)=M_{A,\alpha}$.
\end{proof}

The function $\cM$ is not a bijection in general. 
However this may happen, under the conditions of proposition~\ref{prop:exact}:
this is the \emph{exact parameterization} property from \cite{LPR03},
which is also proved in \cite{DD09}.

\begin{prop}
\label{prop:exact}
With the specifications 
$\Ss_{\eq}$, $\Ss_0$ and $\Ss_A$ as in proposition~\ref{prop:passing},
let $M_0$ be a model of $\Ss_0$ 
and $M_A$ a terminal model of $\Ss_A$ extending $M_0$. 
Then the function $\cM$ from proposition~\ref{prop:passing} is a bijection: 
  $$ M_A(A)  \iso L_{\eq}[\Ss_{\eq},\Set]|_{M_0}  \;.$$
\end{prop}

It follows from \cite{Ru00} and \cite{HR95} that 
there is a terminal model of $\Ss_A$ over $M_0$.
Proposition~\ref{prop:exact} corresponds to the way algebraic structures are implemented 
in the systems Kenzo/EAT.
In these systems 
the parameter set is encoded by means of a record of Common Lisp functions, 
which has a field for each operation in the algebraic structure to be implemented. 
The pure terms correspond to functions which can be obtained 
from the fixed data and do not require an explicit storage. 
Then, each particular instance of the record gives rise to an algebraic structure.



\begin{thebibliography}{}

\bibitem[Barr and Wells 1999]{BW99}
Barr, M. and Wells, C. (1999) Category Theory for Computing Science.
\emph{Centre de Recherches Ma\-th\'e\-ma\-tiques (CRM)
Publications}, 3rd Edition.

\bibitem[Coppey and Lair 1984]{CL84}
Coppey, L. and Lair, C. (1984) Le\c{c}ons de Th\'eorie des
Esquisses. \emph{Diagrammes} \textbf{12}.

\bibitem[Dom\'{\i}nguez and Duval 2009]{DD09} 
Dom\'{\i}nguez, C. and Duval, D. (2009) A parameterization process
as a categorical construction.

\bibitem[Dom\'{\i}nguez et al. 2005]{DDLR05}
Dom\'\i nguez, C., Duval, D., Lamb\'an, L. and Rubio, J. (2005)
Towards diagrammatic specifications of symbolic computation systems.
In: Mathematics, Algorithms, Proofs. Coquand, T., Lombardi, H. and
Roy, M. (Eds.) \emph{Dagstuhl Seminar} \textbf{05021}.
\url{http://drops.dagstuhl.de/portals/index.php?semnr=05021}.

\bibitem[Dom\'{\i}nguez et al. 2007]{DLR07}
Dom\'{\i}nguez, C., Lamb\'an, L. and Rubio, J. (2007)
Object-oriented institutions to specify symbolic computation
systems. \emph{Rairo - Theoretical Informatics and Applications}
\textbf{41} 191--214.

\bibitem[Dom\'{\i}nguez et al. 2006]{DRS06}
Dom\'{\i}nguez, C., Rubio, J. and Sergeraert, F. (2006) Modeling
Inheritance as coercion in the Kenzo system. \emph{Journal of
Universal Computer Science} \textbf{12} (12) 1701--1730.

\bibitem[Dousson et al. 1999]{Kenzo}
Dousson, X., Sergeraert, F. and Siret, Y. (1999) The Kenzo program.
Institut Fourier, Grenoble.
\url{http://www-fourier.ujf-grenoble.fr/~ sergerar/Kenzo}.

\bibitem[Dumas et al. 2009]{DDR09}
Dumas, J.G., Duval, D. and Reynaud, J.C. (2009) Cartesian effect
categories are Freyd-categories. \url{arXiv:0903.3311v3}.

\bibitem[Duval 2003]{Du03}
Duval, D. (2003) Diagrammatic specifications. \emph{Mathematical
Structures in Computer Science} \textbf{13} 857--890.

\bibitem[Duval 2007]{Du07}
Duval, D. (2007) Diagrammatic inference. arXiv:0710.1208v1.

\bibitem[Ehresmann 1968]{Eh68}
Ehresmann, C. (1968) Esquisses et types de structures alg\'ebriques.
\emph{Bull. Instit. Polit. Ia\c{s}i} \textbf{XIV}.

\bibitem[Gabriel and Ulmer 1971]{GU71}
Gabriel, P. and Ulmer, F. (1971)  
\emph{Lokal pr\"asentierbare Kategorien}. Springer
Lecture Notes in Mathematics \textbf{221}. 

\bibitem[Gabriel and Zisman 1967]{GZ67}
Gabriel, P. and Zisman, M. (1967) \emph{Calculus of Fractions and
Homotopy Theory}. Springer.

\bibitem[Goguen and Burstall 1984]{GB84}
Goguen, J. A. and Burstall, R. M. (1984)
\emph{Introducing Institutions}. 
Springer Lecture Notes in Computer Science \textbf{164} 221--256.

\bibitem[Goguen and Malcolm 2000]{GM00}
Goguen, J. and Malcolm, G. (2000) 
A hidden agenda. 
\emph{Theoretical Computer Science} \textbf{245} (1) 55--101.

\bibitem[Hensel and Reichel 1995]{HR95}
Hensel, U. and Reichel, H. (1995) Defining equations in terminal
coalgebras. In: \emph{Recent Trends in Data Type Specifications}, 
Springer Lecture Notes in Computer Science \textbf{906} 307--318.

\bibitem[Kan 1958]{Ka58}
Kan, D.M. (1958) Adjoint Functors.
\emph{Transactions of the American Mathematical Society}
\textbf{87} 294--329.

\bibitem[Lamb\'an et al. 2003]{LPR03}
Lamb\'an, L., Pascual, V. and Rubio, J. (2003) An object-oriented
interpretation of the {E}{A}{T} system. \emph{Applicable Algebra in
Engineering, Communication and Computing} \textbf{14} (3) 187--215.

\bibitem[Lellahi 1989]{Lellahi89}
Lellahi, S.K. (1989) Categorical abstract data type (CADT).
\emph{Diagrammes} \textbf{21}, SKL1-SKL23.

\bibitem[Mac Lane 1998]{MacLane98}
Mac Lane, S. (1998) \emph{Categories for the Working Mathematician}.
Springer, 2th edition.

\bibitem[Makkai 1997]{Makkai97}
Makkai, M. (1997) Generalized sketches as a framework for
completeness theorems (I). \emph{Journal of Pure and Applied
Algebra} \textbf{115} 49--79.

\bibitem[Pitts 2000]{Pitts}
Pitts, A.M. (2000) Categorical Logic. Chapter 2 of Abramsky, S.,
Gabbay, D.M. and Maibaum, T.S.E. (Eds.) \emph{Handbook of Logic in
Computer Science} \textbf{5}. Algebraic and Logical Structures.
Oxford University Press.

\bibitem[Rubio et al. 2007]{EAT}
Rubio, J., Sergeraert, F. and Siret, Y. (2007) EAT: Symbolic
Software for Effective Homology Computation. Institut Fourier,
Grenoble. \url{ftp://fourier.ujf-grenoble.fr/pub/EAT}.

\bibitem[Rutten 2000]{Ru00}
Rutten, J.J.M.M. (2000) Universal coalgebra: a theory of systems.
\emph{Theoretical Computer Science} \textbf{249} (1) 3--80.

\bibitem[Wells 1993]{Wells93}
Wells, C. (1993) Sketches: Outline with References.
\url{http://www.cwru.edu/artsci/math/wells/pub/papers.html}.

\end{thebibliography}
\end{document}